\documentclass[12pt]{article}
\usepackage[centertags]{amsmath}
\usepackage{amsfonts}
\usepackage{amssymb}
\usepackage{amsthm} 
\usepackage{newlfont}
\usepackage{epsfig}
\usepackage{amscd}

\newcommand{\beq}{\begin{equation}}
\newcommand{\eeq}{\end{equation}}
\newcommand{\ba}{\begin{array}}
\newcommand{\ea}{\end{array}}
\newcommand{\bea}{\begin{eqnarray}}
\newcommand{\eea}{\end{eqnarray}}

\begin{document}

\begin{center}
{\large \sc \bf Particular solutions to  multidimensional PDEs with KdV-type nonlinearity. 
}

\vskip 15pt

{\large  A. I. Zenchuk }

\vskip 8pt


\smallskip

{\it  Institute of Chemical Physics, RAS,
Acad. Semenov av., 1
Chernogolovka,
Moscow region
142432,
Russia}

\smallskip

\vskip 5pt

e-mail:  {\tt zenchuk@itp.ac.ru }

\vskip 5pt

{\today}

\end{center}

\begin{abstract}
We consider a class of particular solutions to  the (2+1)-dimensional nonlinear partial differential equation (PDE)  $u_t +\partial_{x_2}^n u_{x_1} - u_{x_1} u =0$ (here $n$ is any integer) reducing it to the ordinary differential equation (ODE). In a 
 simplest  case,   $n=1$, the ODE is solvable in terms of elementary functions. Next choice, $n=2$, yields the cnoidal waves for the special case of Zakharov-Kuznetsov equation. The proposed method is based on the deformation of the characteristic of the equation $u_t-uu_{x_1}=0$ and might also be 
useful in  study the higher dimensional PDEs with arbitrary  linear part and KdV-type nonlinearity  (i.e. the nonlinear term  is $u_{x_1} u$). 
\end{abstract}

\section{Introduction}

Method of characteristics \cite{Whitham} is an effective tool for integrability of first order nonlinear PDEs in arbitrary dimensions. Its modification allowing one to integrate (1+1)-dimensional vector equations (the so called generalized hodograph method) 
was  introduced by Tsarev \cite{T1} and developed in refs.\cite{DN,T2,F}.  A class of matrix equations integrable by the method of characteristics is proposed in \cite{SZ}.

In this paper we develop a method of construction of the particular solutions to a new class of multidimensional PDEs using a deformation of characteristics of already known integrable first-order equations. 
We will refer to  the following first order 
(1+1)-dimensional partial differential equation (PDE) 
\begin{eqnarray}\label{br}
u_t - u_{x_1} u =0,
\end{eqnarray}  which is  a simplest equation  implicitly solvable by the method of  characteristics. It 
describes the break of the wave profile.
It is well known that its general solution is implicitly given  by the non-differential equation
\begin{eqnarray}\label{char}
x+ u t = F(u),
\end{eqnarray}
which is a characteristic of eq.(\ref{br}).
Here $F$ is an arbitrary function which may be fixed by the initial condition.
We show that the properly modified  equation (\ref{char})   allows one to 
partially integrate the  higher-dimensional PDEs, in particular, the (2+1)-dimensional PDE
\begin{eqnarray}\label{pden}
u_t +\partial_{x_2}^n u_{x_1} - u_{x_1} u =0.
\end{eqnarray} 

Note that the term  "partially integrable PDEs"  has two different meanings. First, the PDE is 
partially integrable if the available   solution space is not full (see examples in ref.\cite{ZS}).
Second, the PDE is partially integrable if its  solution space is described in terms of the  
lower dimensional PDEs \cite{Zen}. Our nonlinear  PDEs are partially integrable in both 
senses. More precisely, we consider a family of solutions to the $(2+1)$-dimensional 
nonlinear PDE  (\ref{pden}) which is described by a nonlinear ordinary differential 
equation (ODE) with independent variable $x_2$ where $x_1$ and $t$ appear as parameters. This method may be viewed as a deformation of the method of characteristics \cite{Whitham}.

If $n=1$, then eq.(\ref{pden}) reads
\begin{eqnarray}\label{visc_fl}
u_{t} +u_{x_1 x_2} -  u u_{x_1} =0.
\end{eqnarray}
 In this case we represent the explicite solution parametrized by one
 arbitrary function of single variable. If $n=2$, then eq.(\ref{pde}) yields a special case of Zakharov-Kuznetsov equation \cite{ZK,B,Fam}
 \begin{eqnarray}\label{Z_K}
u_{t}  + u_{x_1x_2x_2} -  u u_{x_1} =0.
\end{eqnarray}
Explicite solutions are  represented in terms of elliptic functions (in particular, in terms of the cnoidal waves) and are parametrized by two arbitrary functions of single variable. 

General algorithm for ($2+1$)-dimensional PDE is represented in the next section, Sec.\ref{Section:general}, together with examples of explicite solutions. The asymptotic solutions at large $t$ are derived as well. Higher dimensional generalization of the  proposed algorithm is discussed in Sec.\ref{Section:ggeneral}. 
Conclusions are given in Sec.
\ref{Section:conclusions}.

\section{General form of  (2+1)-dimensional PDE reducible to ODE}
\label{Section:general}

We proceed with the (2+1)-dimensional nonlinear PDE and prove the following theorem.

{\bf Theorem 1.}
Let function $u$ be a solution of the following ODE (a deformation of eq.(\ref{char}))
\begin{eqnarray}\label{F0}
 x_1 + u t=\frac{1}{2} u^2-\partial_{x_2}^{n}  u
\end{eqnarray}
($n$ is an arbitrary integer) in the domain 
\begin{eqnarray}
\label{str}\label{xx}
a_1  <  x_1 < b_1,\;\;\; a_2 \le x_2 \le b_2,\;\;t>0
\end{eqnarray}
of the space of variables $x_1$, $x_2$
with some boundary conditions
\begin{eqnarray}\label{in}
B^i u\equiv \alpha_{1i} u|_{x_2=a_2} +\alpha_{2i} u|_{x_2=b_2} + 
\alpha_{3i} \partial_{x_2}^{i} u|_{x_2=a_2} +
\alpha_{4i} \partial_{x_2}^{i}u|_{x_2=b_2}  =\chi_i,\;\;i=1,\dots,n-1 ,
\end{eqnarray}
where functions $\chi_i$ are constrained by the linear PDEs
\begin{eqnarray}\label{bpsi2}
(\chi_i)_{t} = (\chi_i)_{x_1} t + \alpha_{1i} + \alpha_{2i},\;\;i=1,\dots,n-1.
\end{eqnarray}
Then the function $u$ is a solution of the nonlinear PDE 
\begin{eqnarray}\label{md}
u_t +\partial_{x_2}^n u_{x_1} -  u u_{x_1}=0,
\end{eqnarray}
with  boundary conditions (\ref{in},\ref{bpsi2}).

{\it Proof.}
To derive eq.(\ref{md}) from eq.(\ref{F0}) we, first,
rewrite  eq.(\ref{F0}) in the form
\begin{eqnarray}\label{F}
x_1 + u t + w =0 ,
\end{eqnarray}
where 
\begin{eqnarray}\label{w}
w(u)= \partial_{x_2}^{n} u -\frac{1}{2} u^2.
\end{eqnarray}
For the further analysis, we shall mention that  eq.(\ref{md})  admits a representation in the form
\begin{eqnarray}\label{uw}
&&
u_t + w_{x_1}(u) =0.
\end{eqnarray}
Moreover, $w$ satisfies equation
\begin{eqnarray}\label{mdw}
w_t +\partial_{x_2}^{n} w_x -  u w_{x_1}=0,
\end{eqnarray}
which may be written as
\begin{eqnarray}\label{Luw}
(\partial_{x_2}^{n}-u)(u_t + w_{x_1}) =0.
\end{eqnarray}
Next, applying the differential 
operators $\partial_t$, $\partial_{x_1}$ and  $L=\partial_{x_1} \partial_{x_2}^n$
to eq.(\ref{F})  we obtain:
\begin{eqnarray}\label{Ft}
E_t&:=& u + t u_t +w_t =0 ,\\\label{Fx} E_x&:=& 1+ t u_{x_1}+w_{x_1} = 0
,\\\label{FL}
E_L&:=&  t \partial_{x_2}^n u_{x_1} + \partial_{x_2}^n w_{x_1} =0 .
\end{eqnarray}
Now let us  consider the combination
\begin{eqnarray}
E_t - E_x u + E_L 
\end{eqnarray}
which reads
\begin{eqnarray}\label{uw0}
t (u_t +\partial_{x_2}^n u_{x_1} -  u u_{x_1}) + w_t +\partial_{x_2}^n  w_{x_1} -  u w_{x_1}=0.
\end{eqnarray}
In virtue of  eqs.(\ref{mdw}) and (\ref{Luw}) we have 
$w_t +\partial_{x_2}^{n} w_x -  u w_{x_1}=(\partial_{x_2}^{n}-u)(u_t + w_{x_1})$, so that 
eq.(\ref{uw0}) can be written in the following form
\begin{eqnarray}\label{uwp}\label{uwpsi}
(t  + \partial_{x_2}^{n}-u ) \psi =0,
\end{eqnarray}
where $\psi$ is defined as follows:
\begin{eqnarray}\label{psi}
\psi=u_t + \partial_{x_2}^n u_{x_1} -  u u_{x_1} .
\end{eqnarray} 
Eq.(\ref{uwpsi}) is the ODE for the function $\psi$.
However, to get 
 original PDE (\ref{md}),
 we need the trivial solution to  eq.(\ref{uwpsi}): $\psi\equiv 0$.
For this purpose we impose the following zero conditions on the  boundary of domain  (\ref{xx}):
\begin{eqnarray}\label{tbpsi}
B^i\psi\equiv 
\alpha_{1i}  \psi|_{x_2=a_2} +  \alpha_{2i} \psi|_{x_2=b_2} + 
\alpha_{3i} \partial_{x_2}^{i}   \psi|_{x_2=a_2} +
\alpha_{4i} \partial_{x_2}^{i} \psi|_{x_2=b_2}  =0,\;\;i=1,\dots,n-1 .
\end{eqnarray}
Now, using the  uniqueness theorem for the linear homogeneous PDE with zero boundary conditions we conclude that the only solution to the boundary-value  problem (\ref{uwpsi},\ref{tbpsi}) is zero, i.e.
$\psi \equiv 0$, which is nothing but eq.(\ref{md}).

We shall   note that  conditions (\ref{tbpsi}) given on the boundary of domain (\ref{xx}) constrain  
the boundary conditions for PDE (\ref{md}).
We derive this constraint assuming that  the boundary conditions for  PDE (\ref{md}) are introduced by the  operators $B^i$  in eq.(\ref{tbpsi}) (i.e. we consider the boundary conditions (\ref{in})). 
For this purpose we apply operators $B^i$ to $E_x$ given by  
eq.(\ref{Fx}) and subtract the result from eq.(\ref{tbpsi}) obtaining 
\begin{eqnarray}\label{B1}
B^i (u_t-u_{x_1} t -1) =0.
\end{eqnarray}
From another hand, applying the operator $\partial_t - t \partial_{x_1}$ to eq.(\ref{in}) 
we obtain
\begin{eqnarray}\label{B2}
B^i (u_t-u_{x_1} t ) = (\chi_i)_t - t (\chi_i)_{x_1}.
\end{eqnarray}
Subtracting eq.(\ref{B2}) from eq.(\ref{B1}) we result in
eq.(\ref{bpsi2}) for  the functions $\chi_i(x_1,t)$, $i=1,\dots,n-1$. 
$\Box$

Below we list several properties of the solution $u$.
\begin{enumerate}
\item
As shown in Theorem 1,
the zero boundary conditions  (\ref{tbpsi}) 
 for the function $\psi$ effect the  boundary conditions for the function $u$ as a solution of the nonlinear PDE (\ref{F0}). Namely, functions $\chi_i$ 
 in boundary  conditions (\ref{in}) are not arbitrary but satisfy  
system of linear eqs.(\ref{bpsi2}). This system  can be readily integrated yielding
\begin{eqnarray}\label{bpsi2_sol}
&&
\chi_i(x_1,t)=t (\alpha_{1i} +\alpha_{2i}) +A_i(\eta),\;\;i=1,\dots,n-1\\\label{eta}
&&
\eta=t^2 + 2 x_1,
\end{eqnarray}
where $A_i$ are arbitrary functions of single argument.
\item
In general, 
the domain (\ref{xx})  of variables $x_1$ and $x_2$ might be either bounded or not bounded, i.e. $a_1$ and $a_2$ might tend to $-\infty$ while $b_1$ and $b_2$ might tend to $+\infty$. 
\item
If the function $u$ is solution to both eqs.(\ref{md}) and (\ref{F0}) then it satisfies the linear (1+1)-dimensional PDE
\begin{eqnarray}\label{pde}
u_t = t u_{x_1}+1.
\end{eqnarray}
To derive eq.(\ref{pde}) we differentiate eq.(\ref{F0}) with respect to $x_1$ and subtract the result  from eq.(\ref{md}). Note that relation (\ref{pde}) on the boundary of domain (\ref{xx}) is forced by relations (\ref{bpsi2}), as follows from the proof of Theorem 1, see eqs.(\ref{B1},\ref{B2}).
\item
In general, eq.(\ref{F0}) is a nonlinear ODE. However, it is remarkable that the asymptotic $u^{as}$ of the bounded solution  in the  domain (\ref{str}) of variables $x_1$ and $x_2$ with the bounded parameters $a_1$ and $a_2$  at large $t$
is described by the linear ODE
\begin{eqnarray}\label{F0as}
  \partial_{x_2}^n u^{as} = -u^{as} t,
\end{eqnarray}
because, in this case, we neglect terms $x_1$ and $\displaystyle \frac{1}{2} u^2$ in eq.(\ref{F0}).
In addition,   we may neglect $x_1$ in  definition of $\eta$ (\ref{eta}) and write
\begin{eqnarray}\label{etaas}
\eta^{as}\equiv t^2.
\end{eqnarray}
Analysis of eq.(\ref{F0as}) depends on the particular $n$ and on the boundary condition.   We consider eq.(\ref{F0as}) in Secs.\ref{Section:neq1} and \ref{Section:neq2} for particular examples of nonlinear PDE (\ref{md}) with $n=1$ and $n=2$.
\end{enumerate}


\subsection{Simplest case $n=1$. Family of explicite solutions of nonlinear PDE }
\label{Section:neq1}
If $n=1$, then 
nonlinear PDE (\ref{md}) reads
\begin{eqnarray}\label{md22ex}
u_t +  u_{x_1x_2} - u u_{x_1} =0.
\end{eqnarray}
In turn, eq.(\ref{F0}) reads
\begin{eqnarray}\label{du2}
u_{x_2} =\frac{1}{2} u^2-x_1 - u t.
\end{eqnarray}
Boundary  condition (\ref{in}) is represented by a  single equation (we take $a_2=0$ without loss of generality) which we take in the simplest form
\begin{eqnarray}\label{bpsiex1}
u|_{x_2=0}=\chi_1(t,x_1).
\end{eqnarray} 
Here $\chi_1$ has the form given in 
eq.(\ref{bpsi2_sol}) with $\alpha_{11}=1$ and $\alpha_{21}=0$: 
\begin{eqnarray}
\label{chi0}
\chi_1(x_1,t) = t + A_1(\eta).
\end{eqnarray}
Eq.(\ref{du2}) is the first order ODE  with separable variables. It may be readily represented in the form  
\begin{eqnarray}\label{duf}
 u_{x_2} =\frac{1}{2}(u-u_+) (u-u_-) ,\;\;u_\pm=t\pm\sqrt{\eta},
\end{eqnarray}
and integrated.
General solution of eq.(\ref{duf}) 
 reads:
\begin{eqnarray}\label{n1u}
u(x_1,x_2,t) &=& 
t + \sqrt{\eta} \frac{ 1+ C(x_1,t)
e^{x_2\sqrt{\eta}}}{ 1- C(x_1,t)
e^{x_2\sqrt{\eta}}}.
\end{eqnarray}
Boundary condition (\ref{bpsiex1}) in virtue of eq.(\ref{chi0}) prescribes the dependence of $C$ only on the variable $\eta$: $C(x_1,t)\equiv C(\eta)$. This condition reads:
\begin{eqnarray}
A_1(\eta)=  \sqrt{\eta} \frac{ 1+ C(\eta)
e^{x_2\sqrt{\eta}}}{ 1- C(\eta)
e^{x_2\sqrt{\eta}}}
\end{eqnarray}
and may be considered as the definition of $A_1(\eta)$ in terms of the  arbitrary function $C(\eta)$.
Thus, eq.(\ref{n1u}) represents a family of solutions of eq.(\ref{md}) parametrized by one function of single  variable:\begin{eqnarray}\label{n1uf}
u(x_1,x_2,t) &=& 
t + \sqrt{\eta} \frac{ 1+ C(\eta)
e^{x_2\sqrt{\eta}}}{ 1- C(\eta)
e^{x_2\sqrt{\eta}}}.
\end{eqnarray}

\subsubsection{Asymptotics of bounded solution}
Consider the problem in the  domain (\ref{xx}) of variables $x_1$, $x_2$ with the finite bounds $a_1$ and  $b_1$
at large $t$. Then we shall use eq.(\ref{etaas}) instead of (\ref{eta}) for $\eta$. Consequently, the asymptotic solution will not depend on $x_1$,  which follows from eq.(\ref{n1uf}) for solution $u$. Let us expand solution  (\ref{n1uf}) in powers of  $e^{-x_2 t}$  and write its  first nontrivial term as 
\begin{eqnarray}\label{n1as}
u=C^{as}(t) e^{-x_2 t}, \;\;C(t) =- \frac{2t}{C^{as}(t)},
\end{eqnarray}
where $C^{as}(t)$ must be a bounded function of $t$.
The same expression may be obtained directly from  asymptotic equation (\ref{F0as}) with $n=1$.
The boundary condition (\ref{bpsiex1})  in virtue of eq.(\ref{chi0}) yields 
\begin{eqnarray}
A_1(t)  = -t +C^{as}(t).
\end{eqnarray}
We see that asymptotic solution (\ref{n1as}) vanishes at $t\to\infty$.

\subsection{Case $n=2$.  Cnoidal waves for a particular case of Zakharov-Kuznetsov equation}
\label{Section:neq2}
If $n=2$, 
then
nonlinear PDE (\ref{md}) reads
\begin{eqnarray}\label{md2}
u_t +  u_{x_1x_2x_2} - u u_{x_1} =0,
\end{eqnarray}
which is a particular case of the Zakharov-Kuznetsov equation \cite{ZK} when the transversal part of the two-dimensional Laplacian dominates (i.e., if the Debye radios in plasma is small in comparison with the $r_{H}=c_s/w_{Hi}$, where $c_s$ is the sound velocity and $w_{Hi}$ is the ion cyclotron frequency in plasma \cite{ZK}).
We choose  the boundary   conditions (\ref{in}) in the form  of two following equations
\begin{eqnarray}\label{bpsiex2}
u|_{x_2=a_2}=\chi_1(t,x_1), \;\;\; u|_{x_2=b_2}=\chi_2(t,x_1),
\end{eqnarray}
where  $\chi_1$, $\chi_2$ are  given by eqs.(\ref{bpsi2_sol}) with
$\alpha_{11}=\alpha_{22}=1$, $\alpha_{12}=\alpha_{21}=0$:
\begin{eqnarray}\label{bpsi20}
\chi_i(t,x_1) =t + A_i(\eta),\;\;i=1,2.
\end{eqnarray}
In our case, eq.(\ref{F0}) reads 
\begin{eqnarray}\label{dwex2}
 u_{x_2x_2} =\frac{1}{2}(u^2 - 2 t u -2 x_1).
\end{eqnarray}
It is equivalent to the following  first-order ODE:
\begin{eqnarray}\label{dwex22}
u_{x_2} =\pm\frac{1}{\sqrt{3}} \sqrt{u^3 -3 t u^2 -6 x_1 u + \tilde C_1(x_1,t)}\,,
\end{eqnarray}
where $\tilde C_1$ is a function to be fixed by the boundary conditions (\ref{bpsiex2}). 
Introducing field $v$ related with $u$ by the equation
\begin{eqnarray}\label{v}
v=u-t,
\end{eqnarray}
we write eq.(\ref{dwex22}) as
\begin{eqnarray}\label{dwex23}
v_{x_2} =\pm \frac{1}{2\sqrt{3}} \sqrt{4 v^3  -12( t^2 +2 x_1) v -4  C_1(x_1,t)},\;\;
 C_1(x_1,t) =6 t x_1+2 t^3 -
\tilde C_1(x_1,t).
\end{eqnarray}
The solution of eq.(\ref{dwex23})  is  the elliptic Weierstrass function with invariants $g_2=12( t^2 +2 x_1)$ and $g_3(C_1(x_1,t))=4  C_1(x_1,t)$:
\begin{eqnarray}\label{Ww}
v=u-t={\cal{W}}\left(\pm \frac{x_2}{2\sqrt{3}} +  C_2(x_1,t) ,g_2,g_3( C_1(x_1,t))\right) .
\end{eqnarray}
The functions $ C_i$, $i=1,2$ may be found from boundary conditions 
(\ref{bpsiex2}). Putting successively
 $x_2=a_2$ and $x_2=b_2$  in 
eq.(\ref{Ww}) and using eq.(\ref{bpsi20}) we see that $C_i$ must depend only 
on $\eta$.
Thus, eq.(\ref{Ww})  represents  solution of eq.(\ref{md2}) parametrized by two arbitrary  functions of single variable:
\begin{eqnarray}\label{Wwu}
u=t+{\cal{W}}\left(\pm \frac{x_2}{2\sqrt{3}} +  C_2(\eta) ,g_2,g_3( C_1(\eta))\right) .
\end{eqnarray}
Boundary conditions (\ref{bpsiex2}) in virtue of eqs.(\ref{bpsi20}) read 
\begin{eqnarray}\label{Ww2}
&&
A_1(\eta)={\cal{W}}\left(\pm \frac{a_2}{2\sqrt{3}} C_2(\eta) ,g_2,g_3(C_1(\eta))\right) ,\\\nonumber
&&
A_2(\eta)={\cal{W}}\left(\pm \frac{b_2}{2\sqrt{3}} C_2(\eta) ,g_2,g_3(C_1(\eta))\right) .
\end{eqnarray}
which must  be taken as the definitions of the functions $A_i(\eta)$ in terms of the arbitrary functions $C_i(\eta)$, $i=1,2$.

Now we consider a particular case  when 
all the  roots of the polynomial  under the square root in the rhs of eq.(\ref{dwex23}), i.e the roots of the polynomial
\begin{eqnarray}\label{pol}
 v^3  -3( t^2 +2 x_1) v - C_1(x_1,t) =0,
\end{eqnarray}
 are real.
These roots read
\begin{eqnarray}\label{ppp}
&&
p_1=d + \frac{\eta}{d},\\\nonumber
&&
p_2=\frac{1}{2} d (-1+i\sqrt{3})  -\frac{1}{2 d} (1+i \sqrt{3})\eta = -\frac{p_1}{2}+\frac{\sqrt{3 }i}{2}\left(d-\frac{\eta}{d}\right),\\\nonumber
&&
p_3=\frac{1}{2} d (-1-i\sqrt{3})  -\frac{1}{2 d} (1-i \sqrt{3})\eta = -\frac{p_1}{2}-\frac{\sqrt{3 } i}{2}\left(d-\frac{\eta}{d}\right) =-p_1-p_2,
\end{eqnarray}
where
\begin{eqnarray}\label{d}
d=2^{-1/3}\left(  C_1(x_1,t) +\sqrt{ C_1^2(x_1,t) -4 \eta^3}
\right)^{1/3}.\;\;\;
\end{eqnarray}
Thus, all $p_i$ are parametrized by one  function  $ C_1(x_1,t)$ which will be fixed by the boundary conditions.
These roots are real if $|d|^2=\eta$, which is the identity provided  $ C_1^2(x_1,t)-4 \eta^3 \le 0$. 
In this case, 
let 
\begin{eqnarray}\label{minmaxp}
v_1(x_1,t)=\min(p_1,p_2,p_3),\;\; 
v_3(x_1,t)=\max(p_1,p_2,p_3),
\end{eqnarray}
and $v_2(x_1,t)$ be the remaining root out of the list $(p_1,p_2,p_3)$, i.e.
$v_1\le v_2\le v_3$. Now we write   eq.(\ref{dwex23}) as
\begin{eqnarray}\label{dwex24}
v_{x_2}=\pm \frac{1}{\sqrt{3}}\sqrt{(v-v_1)(v-v_2) (v-v_3)}.
\end{eqnarray}
Let us assume that $v$ is bounded as
\begin{eqnarray}
v_1\le v \le v_2.
\end{eqnarray}
Then
the substitution
\begin{eqnarray}
q=\sqrt{\frac{v-v_1}{v_2-v_1}},\;\;0\le q \le 1
\end{eqnarray}
transforms eq.(\ref{dwex24}) into the following form
\begin{eqnarray}\label{dwex25}
&&
q_{x_2}=\pm \frac{1}{2}\sqrt{\frac{v_3-v_1}{3}}\sqrt{(1-q^2)(1-k^2 q^2)},\\\label{k}
&&
k^2=\frac{v_2-v_1}{ v_3-v_1}.
\end{eqnarray}
Solution to this equation is the  Jacoby elliptic function
\begin{eqnarray}\label{q}
q=\sqrt{\frac{u-t-v_1}{ v_2-v_1}} = {\mbox{sn}}\Big(
\pm
\frac{x_2\sqrt{v_3-v_1}}{2\sqrt{3}}+\tilde C_2(x_1,t);k\Big) .
\end{eqnarray}
Solving eq.(\ref{q}) for $u$, we obtain
\begin{eqnarray}\label{uu0}
u=t+ v_1(\eta) + (v_2(x_1,t)-v_1(x_1,t)){\mbox{sn}}^2\Big(\pm
\frac{x_2\sqrt{v_3(x_1,t)-v_1(x_1,t)}}{2\sqrt{3}}+\tilde C_2(x_1,t);k\Big).
\end{eqnarray}
Next, boundary conditions (\ref{uu}) suggest us to take $C_1$ and $\tilde C_2$ as functions of $\eta$,
$ C_1(x_1,t) \equiv C_1(\eta)$, $ \tilde C_2(x_1,t) \equiv \tilde C_2(\eta)$. This means that $v_i(x_1,t) \equiv v_i(\eta)$, $i=1,2,3$. In addition, 
 ${\mbox{sn}}^2(x)$ is an even function of argument. Consequently, we may  rewrite eq.(\ref{uu0}) as (remember that $v_i(\eta)$, $i=1,2,3$, depend on $C_1(\eta)$ through eqs.(\ref{ppp},\ref{d}))
\begin{eqnarray}\label{uu}
u=t+ v_1(\eta)+ (v_2(\eta)-v_1(\eta)){\mbox{sn}}^2\Big(
\frac{x_2\sqrt{v_3(\eta)-v_1(\eta)}}{2\sqrt{3}}+ C_2(\eta);k\Big),\;\;  C_2(\eta) =\pm
\tilde C_2(\eta).
\end{eqnarray}
Boundary conditions (\ref{bpsiex2}) in virtue of eqs.(\ref{bpsi20}) yield 
\begin{eqnarray}\label{uuw}
&&
A_1(\eta)= v_1(\eta) + (v_2(\eta)-v_1(\eta))\,{\mbox{sn}}^2\Big(
\frac{a_2\sqrt{v_3(\eta)-v_1(\eta)}}{2\sqrt{3}}+C_2(\eta);k\Big) ,\\\nonumber
&&
A_2(\eta)= v_1(\eta) + (v_2(\eta)-v_1(\eta))\,{\mbox{sn}}^2\Big(
\frac{b_2\sqrt{v_3(\eta)-v_1(\eta)}}{2\sqrt{3}}+C_2(\eta);k\Big) ,
\end{eqnarray}
which must be viewed as the definitions of $A_i$ in terms of the arbitrary functions $ C_i$, $i=1,2$. 
Consequently, eq.(\ref{uu}) represents a family of solutions of eq.(\ref{md2})  with two arbitrary functions $C_i$, $i=1,2$, of single variable $\eta$.

\subsubsection{Asymptotic of bounded solutions}
We consider the bounded solution on the domain (\ref{xx}) of variables $x_1$ and $x_2$ with finite bounds $a_1$ and $b_1$ at large $t$. 
Asymptotic eq.(\ref{F0as}) reads:
\begin{eqnarray}\label{dwex2as}
 u^{as}_{x_2x_2} =- t u^{as} .
\end{eqnarray}
This equation  has the following general solution
\begin{eqnarray}\label{uC}
u^{as}=C_1^{as}(x_1,t) \cos\Big(\sqrt{t} x_2  +C_2^{as}(x_1,t) \Big),
\end{eqnarray}
where $C_1^{as}$ is an amplitude and $C_2^{as}$ is a phase. 
We consider $C_1^{as}(x_1,t)\ge 0$ without loss of generality (sign "$-$" may be embedded in $C_2^{as} $). 
At large $t$, we use eq.(\ref{etaas}) instead of (\ref{eta}) for $\eta$. Then $A_i$, $i=1,2$, may be considered as functions of  $t$. The boundary functions $\chi_i(t)$  (\ref{bpsi20}) become bounded if
\begin{eqnarray}\label{ABas}
A_i(t)=-t  +B_i(t),\;\;i=1,2,
\end{eqnarray}
where $B_i(t)$, $i=1,2$, are bounded functions of argument. Boundary conditions (\ref{bpsiex2}) at $t\to\infty$  suggest us to take 
$C_i^{as}$ as functions of $t$, so that eq.(\ref{uC}) reads
\begin{eqnarray}\label{uC2}
u^{as}=C_1^{as}(t) \cos\Big(\sqrt{t} x_2  +C_2^{as}(t) \Big).
\end{eqnarray}
Thus, asymptotic (\ref{uC2}) does not depend on $x_1$
and is parametrized by two arbitrary functions of $t$, $C^{as}_i(t)$, $i=1,2$.
In the asymptotic case,  boundary conditions (\ref{bpsiex2}) in virtue of eqs.(\ref{bpsi20})  read
\begin{eqnarray}\label{C01}
B_1(t)=C^{as}_1(x_1,t) \cos\Big(a_2\sqrt{t}   +C^{as}_2(t) \Big),\;\;
B_2(t)=C^{as}_1(x_1,t) \cos\Big(b_2\sqrt{t}   +C^{as}_2(t) \Big).
\end{eqnarray}
which must be taken as definitions of $B_i(t)$ in terms of  the arbitrary functions $C_i(t)$, $i=1,2$.

\paragraph{Asymptotics of cnoidal waves (\ref{uu}).}
Now we show that  asymptotic solution (\ref{uC2}) is nothing but the  asymptotic of cnoidal wave (\ref{uu}). 

Since $\eta$ is given by eq.(\ref{etaas}) at large $t$, we neglect $x_1$-dependence in $C_i$ and take $ C_1(t) =2 t^3 -  t \hat C_1(t)$ with bounded $\hat C_1(t)\ge 0$. Then eq.(\ref{d}) yields
$d^{as}=t + \frac{i \sqrt{\hat C_1}}{3}$. Next we  write the roots $p^{as}_i\equiv p_i|_{t\to\infty}$, given by eqs.(\ref{ppp}), as follows:
\begin{eqnarray}
p^{as}_1 = 2t,\;\;p_2^{as}=-t -\sqrt{\frac{\hat C_1}{3}},\;\;p^{as}_3=-t +\sqrt{\frac{\hat C_1}{3}}.
\end{eqnarray}
Thus,  in the asymptotic limit,  we have explicite map  between $v^{as}_i$ and $p^{as}_i$,
\begin{eqnarray}\label{minmaxp2}
v^{as}_1=p^{as}_2= -t -\sqrt{\frac{\hat C_1}{3}}, \;\;\;v^{as}_2=p^{as}_3= -t +\sqrt{\frac{\hat C_1}{3}},\;\;\;
v^{as}_3=p^{as}_1= 2 t,
\end{eqnarray}
unlike the non-asymptotic case, see eqs.(\ref{minmaxp}).
Now we may calculate
\begin{eqnarray}\label{uas}
&&
v^{as}_2-v^{as}_1 =2\sqrt{\frac{\hat C_1}{3}},\;\;\;
v^{as}_3-v^{as}_1 =3t +\sqrt{\frac{\hat C_1}{3}}.
\end{eqnarray}
Consequently,  for $k^2$ in eq.(\ref{k}) we have
\begin{eqnarray}
k^2=\frac{2\sqrt{\frac{\hat C_1}{3}}}{3t +\sqrt{\frac{\hat C_1}{3}}} \to 0,
\;\;\;{\mbox{at}} \;\;\;t\to\infty. 
\end{eqnarray} 
Since ${\mbox{sn}}(x;k)|_{k=0}=\sin(x)$,
and $\sin^2 x=\frac{1}{2} (1-\cos 2 x)$, we reduce eq.(\ref{uu}) to eq.(\ref{uC2}) where 
\begin{eqnarray}
C_1^{as}(t)=-\sqrt{\frac{\hat C_1}{3}},\;\;
C_2^{as}(t)=  2  C_2(t).
\end{eqnarray}
In turn, conditions (\ref{uuw}) result in eqs.(\ref{C01}) provided eqs.(\ref{ABas}).

\section{Generalization of algorithm to higher dimensions}
\label{Section:ggeneral}
The algorithm developed in Sec.\ref{Section:general} may be generalized to higher dimension.
In this section we introduce a large manifold of $(M+1)$-dimensional nonlinear PDEs  with particular solutions satisfying either $M$- or $(M-1)$-dimensional nonlinear PDEs where the independent variables  $t$ (and sometimes $x_1$) appears as parameter. 
The increase in the dimensionality of the original  nonlinear PDE is achieved  through the increase in the dimensionality of its linear part, while the nonlinearity remains the same. 

We shall remark that one could apply this algorithm to larger class of evolution PDEs with integro-differential linear part. This would be also important because 
 this structure has, in particular, the multidimensional dispersion-less KP. In this respect, we have to remember ref.\cite{MS}, where another type of 
particular solutions to this equation were studied and the wave breaking for such solutions was explained. However, multidimensional evolution PDEs with integro-differential linear part are out of the scope of this paper.

We formulate the  following theorem.

{\bf Theorem 2.}
Let the differential operator $L$ have the  structure 
 $L=\partial_{x_1} \tilde L$, where 
$\tilde L(\partial_{x_i},\;i=1,2 \dots,M)$ is an  arbitrary $M$-dimensional linear differential operator.
Let $u$ be a solution of the
$M$-dimensional PDE
\begin{eqnarray}\label{gF0}
  x_1 + u t=\frac{1}{2} u^2-\tilde L  u
\end{eqnarray}
in the  $M$-dimensional domain of the  space of the variables $x_i$, $i=1,\dots,M$,
\begin{eqnarray}\label{gxx}
a_i\le x_i\le b_i,\;\;i=1,\dots,M,\;\;t>0
\end{eqnarray}
with the complete set of boundary conditions 
\begin{eqnarray}\label{gu}
B^iu =\chi_i, \;\;i=1,\dots,K,
\end{eqnarray}
where $K$ is some integer and functions $\chi_i$, $i=1,\dots,K$, satisfy constraints
\begin{eqnarray}\label{gconstr}
(\chi_i)_t - (\chi_i)_{x_1} =B^i 1,\;\;i=1,\dots,K.
\end{eqnarray}
 Then the function $u$ is a solution of the $(M+1)$-dimensional PDE
\begin{eqnarray}\label{gmd}
u_t +L u -  u u_{x_1}=0.
\end{eqnarray} 

{\it Proof.}
To derive eq.(\ref{gmd}) from eq.(\ref{gF0}) we, first, write eq.(\ref{gF0}) as follows:
\begin{eqnarray}\label{gF}
x_1 + u t + w =0 ,
\end{eqnarray}
where 
\begin{eqnarray}
\label{gw}
w=\tilde  L  u -\frac{1}{2} u^2.
\end{eqnarray}
Similar to eq.(\ref{md}), 
 eq.(\ref{gmd}) admits a  representation in the form
\begin{eqnarray}\label{ggmd}
&&
u_t + w_{x_1} =0, 
\end{eqnarray}
which is the $M$-dimensional PDE.
Again, $w$ satisfies the equation
\begin{eqnarray}\label{gmdw}
w_t +L w -  u w_{x_1}=0,
\end{eqnarray}
which may be written as
\begin{eqnarray}\label{gLuw}
(\tilde L-u)(u_t + w_{x_1}) =0.
\end{eqnarray}
Next, we apply operators $\partial_{t}$, $\partial_{x_1}$ and $L$ to eq.(\ref{gF}) obtaining:
\begin{eqnarray}\label{gFt}
E_t&:=& u + t u_t +w_t =0 ,\\\label{gFx}
E_x&:=& 1+ t u_{x_1}+w_{x_1} = 0,\\
\label{gFL}
E_L&:=&  t L u + L w =0 .
\end{eqnarray}
Now we consider the following  combination
\begin{eqnarray}
E_t - E_x u + E_L 
\end{eqnarray}
which reads
\begin{eqnarray}\label{guw0}
t (u_t +L u -  u u_{x_1}) + w_t +L w -  u w_{x_1}=0.
\end{eqnarray}
In virtue of eqs.(\ref{gmdw}) and (\ref{gLuw}) we may write
$w_t +L w -  u w_{x_1} =(\tilde L-u)(u_t + w_{x_1})$.
Then eq.(\ref{guw0}) gets the following form,
\begin{eqnarray}\label{guw}\label{guwpsi}
(t  + \tilde L - u ) \psi =0,
\end{eqnarray}
where $\psi$ is defined as follows:
\begin{eqnarray}\label{gpsi}
\psi=u_t +L u -  u u_{x_1} .
\end{eqnarray} 
Eq.(\ref{guwpsi}) is the $M$-dimensional PDE  for the function $\psi$.
Let us impose the zero boundary conditions for the function $\psi$
\begin{eqnarray}\label{gB}
B^i\psi =0, \;\;i=1,2,\dots, K.
\end{eqnarray}
Then, in virtue of the uniqueness theorem for the linear homogeneous PDEs, we conclude that $\psi\equiv 0$, which coincides with 
 PDE (\ref{gmd}).
 
Boundary conditions (\ref{gB}) lead to constraints (\ref{gconstr}) for the
 boundary functions $\chi_i$, $i=1,\dots,K$. These constraints may be derived similarly to the derivation of eqs.(\ref{bpsi2}). First, we apply operators $B^i$ to  eq.(\ref{gFx}):
 \begin{eqnarray}
 B^i E_x =0,\;\;i=1,2,\dots, K.
 \end{eqnarray}
  Subtracting this equation from the eq.(\ref{gB}) we obtain
  \begin{eqnarray}\label{gB1}
  B^{(i)}(u_t - t u_{x_1}-1)=0,\;\;i=1,2,\dots, K.
  \end{eqnarray}
  From another hand, applying the operator $\partial_t - t \partial_{x_1}$ to eq.(\ref{gu}) 
we obtain
\begin{eqnarray}\label{gB2}
B^i (u_t-u_{x_1} t ) = (\chi_i)_t - t (\chi_i)_{x_1},\;\;i=1,2,\dots, K.
\end{eqnarray}
Finally, subtracting eq.(\ref{gB2}) from eq.(\ref{gB1}) we derive
eq.(\ref{gconstr}) for  the functions $\chi_i(x_1,t)$, $i=1,\dots,K$. 
$\Box$

Function $u$ possesses the properties  similar to those  of  the function $u$ in    Sec.\ref{Section:general}.
\begin{enumerate}
\item
In general, eq.(\ref{gF0}) is the $M$-dimensional PDE. However, if the operator $\tilde L= \tilde L(\partial_{x_2},\dots,\partial_{M}) $ (i.e. $\tilde L$ is the $(M-1)$-dimensional operator),
then eq.(\ref{gF0}) becomes the $(M-1)$-dimensional PDE. Therewith the variable $x_1$ appears as a parameter in this equation.
\item
The zero boundary conditions (\ref{gB})
 for the function $\psi$ effect the  boundary conditions (\ref{gu}) for the function $u$ through the liner system of (1+1)-dimensional PDEs (\ref{gconstr}) for the functions $\chi_i$, $i=1,\dots,K$.
\item
In general, 
the domain (\ref{gxx})  of variables $x_i$, $i=1,\dots,M$,  might be either bounded or not bounded, i.e. $a_i$ might tend to $-\infty$ while $b_i$ might tend to $\infty$. 
 \item
If the function $u$ is a solution to both eqs.(\ref{gmd}) and (\ref{gF0}) then  it satisfies the linear (1+1)-dimensional PDE (\ref{pde}).
To derive this equation we differentiate eq.(\ref{gF0}) with respect to $x_1$ and subtract the result  from eq.(\ref{gmd}). Note that  this property on the boundary of domain (\ref{gxx})  is forced by relations (\ref{gconstr}) as follows from   the proof of Theorem 2, see eqs.(\ref{gB1},\ref{gB2}).
\item
In general, eq.(\ref{gF0}) is a nonlinear ODE. However, it is remarkable that the asymptotic $u^{as}$ of the bounded solution in the  domain (\ref{gxx}) of variables $x_i$, $i=1,\dots,M$,  with bounded parameters $a_1$ and $b_1$  at large $t$
is described by the linear PDE
\begin{eqnarray}\label{gF0as}
 \tilde L  u^{as} = -u^{as} t,
\end{eqnarray}
because, in this case, we neglect terms $x_1$ and $\displaystyle\frac{1}{2} u^2$ in eq.(\ref{gF0}). We also must use expression (\ref{etaas}) instead of (\ref{eta}) for $\eta$.

Analysis of this equation depends on the particular $\tilde L$ and will  not be considered in
this paper.
\end{enumerate}

\section{Conclusions}
\label{Section:conclusions}

We consider a method of construction of the particular solutions to nonlinear PDE (\ref{md}). In the particular cases, $n=1,2$, these solutions are expressed in terms of 
the  either  elementary functions with  one arbitrary function of single variable ($n=1$) or elliptic functions with two arbitrary functions of single variable  ($n=2$).  The
Zakharov-Kuznetsov equation describing the ion-sound waves in a low-pressure magnetized  plasma  is the physically applicable example of such systems \cite{ZK}. 
We construct the cnoidal waves for the
special case of the Zakharov-Kuzntesov equation, when the two-dimensional Laplacian becomes the one-dimensional transversal one. Asymptotics of these waves  in the bounded space-domain are described by a linear PDE yielding oscillating behavior.  

Generalization of this method to higher dimensions is  briefly discussed. 
In general, a nonlinear $(M+1)$-dimensional  PDE from the considered class  (\ref{gmd}) possesses the family of solutions  satisfying either  $M$- or $(M-1)$-dimensional differential equation (\ref{gF0}). Although we consider only evolutionary type  differential equations, a certain type  of the evolutionary  equations with 
integro-differential linear part might be also studied in this way.

Emphasize that we use deformation of  characteristic (\ref{char}) in our constructions. However, 
an open problem is whether deformations of characteristics  of 
other  multidimensional PDEs might be used in a similar way.

The author thanks Prof. E.A.Kuznetsov for the motivation of this work. 
This work is supported by 
the Program for Support of Leading Scientific Schools 
(grant No. 6170.2012.2).


\end{document}